\documentclass[twocolumn,showpacs,aps]{revtex4}
\usepackage{epsfig}
\usepackage{amsmath}

\begin{document}

\title{A First-Principles Approach to Insulators in Finite Electric Fields}

\author{Ivo Souza, Jorge \'I\~niguez, and David Vanderbilt}

\affiliation{Department of Physics and Astronomy, Rutgers
University, Piscataway, New Jersey 08854-8019}

\begin{abstract}

We describe a method for computing the 
response of an insulator to a 
static, homogeneous electric field.
It consists of iteratively 
minimizing an electric enthalpy functional 
expressed in terms of occupied 
Bloch-like states on a uniform grid of $k$ points. The functional has 
equivalent 
local minima below a critical field ${\cal E}_{\rm c}$ that depends
inversely on the density of $k$ points; the disappearance of the minima at
${\cal E}_{\rm c}$ signals the onset of 
Zener breakdown.
We illustrate the procedure by computing
the piezoelectric and
nonlinear dielectric susceptibility tensors of III-V semiconductors.

\end{abstract}
\date{May 29, 2002}
\pacs{PACS: 77.22.Ch, 78.20.Bh, 42.70.Mp}

\maketitle

The response of insulators and semiconductors to external electric 
fields is of fundamental as well as practical interest. 
It determines their dielectric, piezoelectric, and ferroelectric behavior. 
Much current technological interest is focused on the use of
static fields to tune properties
such as the dielectric function in the RF and microwave region or
the index of 
refraction in the optical region.
Although a sophisticated physical understanding of electric field
effects has emerged~\cite{kane59,wannier60,nenciu91}, 
for the most part this has not translated into tractable computational schemes 
applicable to periodic solids.  Efficient first-principles methods 
for computing derivatives of the total energy of solids with 
respect to a macroscopic field $\boldsymbol{\cal E}$ at $\boldsymbol{\cal E}=0$
do exist~\cite{gironcoli89,dalcorso96,putrino00,nunes01}.
While for many applications such perturbation approaches are adequate,
their extension to nonlinear order is awkward, and for some studies
(e.g., field-induced structural phase transitions~\cite{sai02})
it is essential to perform calculations directly at {\it finite} fields.

The main difficulty is the nature of the 
scalar potential ``$-\boldsymbol{\cal E}\cdot{\bf r}$'', which is 
{\it nonperiodic} and {\it unbounded from below}.
The fact that it destroys the periodicity of the crystal potential
means that methods based on Bloch's theorem do not apply.
Moreover, as a result of 
the unbounded nature of the perturbation, the energy can always be
lowered by transferring charge from the valence states in one region to
conduction states in a distant region.  This is the intrinsic dielectric 
breakdown caused by interband (or Zener) 
tunneling~\cite{kane59,wannier60,nenciu91,zener34}.
In many practical situations Zener tunneling is negligible on the
relevant time scale, and for relatively small fields the system
remains in a polarized long-lived resonant state.
This is the state we would like to describe.
However, the absence of a well-defined ground state invalidates the
variational principle that underlies the usual time-independent
electronic-structure methods and leads to problematic ``runaway''
solutions in implementations of such approaches~\cite{foot-atoms}.

The few first-principles methods that have been proposed for dealing with 
finite fields in solids have had limited success. 
The supercell ``sawtooth'' approach~\cite{kunc83} becomes prohibitively 
expensive for all but the simplest systems. A significant advance,
rooted on the ``modern theory of polarization''~\cite{ksv93},
was the development of a real-space
method based on truncated field-polarized Wannier functions~\cite{nunes94}, 
which removed the need for supercells; however,
the full first-principles implementation~\cite{fernandez98}
was hindered by convergence problems and proved too cumbersome to
find widespread use.

In this Letter we propose an alternative variational approach.
It is based on the minimization of an electric enthalpy functional ${\cal F}$
with respect to Bloch-like functions, where ${\cal F}$ is comprised
of the usual Kohn-Sham energy $E_{\rm KS}$ and a field coupling term
$-{\bf P}_{\rm mac}\cdot\boldsymbol{\cal E}$. Here $E_{\rm KS}$
and the polarization ${\bf P}_{\rm mac}$ are expressed
in terms of a set of field-polarized Bloch functions, the latter via
the Berry-phase theory of polarization~\cite{ksv93}.  Although for
$\boldsymbol{\cal E}\not=0$ the Bloch functions are not eigenstates, they form
an appropriate
representation of the electronic system.  We justify this approach,
showing that a suitable choice of
Brillouin zone sampling prevents
runaway solutions.  We demonstrate its implementation into a standard
electronic band structure program by computing
piezoelectric and linear and nonlinear 
dielectric properties of III-V 
semiconductors,
finding good agreement with experiment. 
Our method opens up new possibilities for first-principles 
investigation of electric-field effects in condensed matter.

We solve for field-polarized Bloch functions 
$\psi_{n \bf k}({\bf r})=e^{i{\bf k}\cdot{\bf r}}\, u_{n \bf k}({\bf r})$
[where $u_{n \bf k}({\bf r})=u_{n \bf k}({\bf r+R})$]
by minimizing the electric enthalpy functional introduced in 
Ref.~\cite{nunes01},
\begin{equation}
\label{enthalpy}
{\cal F}[u_{n \bf k};\boldsymbol{\cal E}]=
E_{\rm KS}- \Omega\, {\bf P}_{\rm mac} \cdot \boldsymbol{\cal E},
\end{equation}
where $\Omega$ is the primitive cell volume and
${\bf P}_{\rm mac}={\bf P}_{\rm ion}+ {\bf P}_{\rm el}$ is the
macroscopic polarization. In a continuous-$k$ formulation,
${\bf P}_{\rm el}$ is $-fe/(2\pi)^3$ times
the sum of valence-band Berry phases~\cite{ksv93}
$\int_{\rm BZ} d{\bf k}\,
\langle u_{n{\bf k}} | \, i\nabla_{\bf k} \, | u_{n{\bf k}} \rangle$
($f$ is the spin degeneracy and $e>0$).
However, as we show below, it is essential to use 
a formulation in terms of a mesh of
$N_k=N_1 \times N_2 \times N_3$ $k$ points in the Brillouin zone (BZ).
Then $ E_{\rm KS}=(f/N_k) \sum_{nj} \langle u_{n{\bf k}_j}
|\, \hat{H}_{\rm KS}({{\bf k}_j}) | u_{n{\bf k}_j} \rangle$ and
${\bf P}_{\rm el} \cdot {\bf b}_i = (fe/\Omega)\,\varphi^{(i)}_{\rm el}$
where
\begin{equation}
\label{berry-phase}
\varphi^{(i)}_{\rm el}=\frac{1}{N_{\perp}^{(i)}}\, 
\sum_{l=1}^{N_{\perp}^{(i)}}\,{\rm Im}\, 
\ln \prod_{j=0}^{N_i-1}\, \det\, 
S\big({\bf k}_{j}^{(i)},
{\bf k}_{j+1}^{(i)}\big)
\end{equation}
is the string-averaged {\it discretized} Berry phase along the direction of 
primitive reciprocal lattice vector ${\bf b}_i$~\cite{ksv93}.
Here $S_{nm}({\bf k},{\bf k}')=
\langle u_{n{\bf k}} | u_{m{\bf k}'} \rangle$,
$n$ and $m$ run over the $M$ occupied bands,
and $N_{\perp}^{(1)}=N_2 \times N_3$ is the number
of strings along ${\bf b}_1$, each containing
$N_1$ points ${\bf k}_j^{(1)}={\bf k}_{\perp}^{(l)}+(j/N_1){\bf b}_1$.

It is implicit in this formulation that when $\boldsymbol{\cal E}\ne0$ the 
electronic 
structure can continue to be represented in terms of field-polarized Bloch-like
functions $\psi_{n{\bf k}}$, {\it even though they are no longer eigenstates of
the Hamiltonian.}
It is well known that when describing an occupied subspace,
one has the freedom to carry out an arbitrary unitary transformation among
the states used to represent it.  In this spirit,
we assume that the density matrix can be written as
$\rho({\bf r},{\bf r}') =(1/N_k) \sum_{n{\bf k}}
\psi^*_{n{\bf k}}({\bf r}') \psi_{n{\bf k}}({\bf r})$,
where $n$ runs over the same number $M$ of Bloch-like states at all $\bf k$.
Then $\rho({\bf r},{\bf r}')=\rho({\bf r+R},{\bf r'+R})$ and
it follows that the charge density and other local quantities are
periodic under translation by a lattice vector $\bf R$.
These are familiar properties of an insulating ground state, and
they turn out to hold for the field-polarized state as well, since:
(i) If one starts with an insulating ground state and applies a homogeneous 
electric field with arbitrary time dependence, the occupied manifold preserves
the above ``insulating-like'' properties at later times. (ii) A state that 
minimizes ${\cal F}$ is a
stationary solution  to the time-dependent Schr\"odinger equation 
in the presence of a static field; since it can be obtained by adiabatically 
turning on the field, it is guaranteed by (i)
to have the above ``insulating-like'' 
properties.
Proofs of (i-ii) are not difficult; details will be given 
elsewhere~\cite{unpublished}.

Usually, the transition to the discrete $k$ space is introduced for merely 
computational reasons. Here, on the contrary, {\it the discrete $k$ formulation
is required to eliminate the possibility of runaway solutions}, i.e., to
allow for stable stationary 
solutions to exist, unaffected by Zener charge-leakage.
This is consistent with previous work showing that in the 
continuous-$k$ limit there are no stationary solutions to the static electric 
field problem
(for an overview, see Sec.~II.\,D of Ref.~\cite{nenciu91}).
To understand why the discretization procedure endows 
${\cal F}$ with minima, it is useful to think of it as 
``bending'' space into a {\it finite} ring:
a uniform mesh of $N_1 \times N_2 \times N_3$ $k$ points
amounts to imposing periodic boundary conditions --~which have a ring 
topology~-- over a supercell of dimensions $L_i=N_i a_i$ ($i$=1,2,3).
For a given $k$-point mesh, ${\cal F}$ will have
minima only if $\boldsymbol{\cal E}$ is small enough that
Zener tunneling is effectively suppressed. This should happen as
long as the distance across which the electrons would have to tunnel in order 
to lower their energy is larger than the ring perimeter $L_i=N_i a_i$. By 
thinking of the field as spatially tilting the energy bands,
one arrives at the condition 
$\boldsymbol{\cal E}\cdot{\bf a}_i<\boldsymbol{\cal E}_{\rm c}\cdot{\bf a}_i$
where $e\boldsymbol{\cal E}_{\rm c}\cdot{\bf a}_i \simeq E_{\rm gap}/N_i$ and
$E_{\rm gap}$ is a representative energy gap.
We have confirmed 
this behavior in a one-dimensional three-band tight-binding 
model~\cite{nunes94} by studying the stability of the field-polarized
solutions and by checking that, for a given $k$ point mesh, 
$E_{\rm gap}/eaN_k$ is a reasonably good estimate of 
${\cal E}_c$.

The polarized state below $\boldsymbol{\cal E}_c$ has additional 
insulating-like 
properties, namely the absence of a steady-state current and the localization
of the Wannier functions to a small portion of the ring~\cite{kohn68,souza00}. 
This state is related to the zero-field ground state by a continuous 
``deformation.'' Such ``polarized manifolds'' have been discussed in the
literature~\cite{kane59,wannier60,nenciu91} for infinite crystals, where they 
were characterized as long-lived resonances. 
Instead, for our finite ring the state obtained by minimizing ${\cal F}$ is 
truly stationary, as discussed above.

By virtue of the nature of the Berry phase term,
the functional ${\cal F}$ cannot be recast as the 
expectation value of a Hermitian operator. Because that term contains  
overlaps between the states at neighboring $k$ points, even in a tight-binding
model without charge self-consistency, the problem must be
solved self-consistently among all $k$ points.
This breakdown of Bloch's theorem is the price to pay
for handling, within periodic boundary
conditions, a field whose scalar potential breaks translational invariance.
Indeed, the Berry-phase term in ${\cal F}$ is the proper
replacement of the usual scalar potential
term $e\boldsymbol{\cal E}\cdot \langle {\bf r} \rangle$ when
using a ring topology.
(Alternatively, one can switch to a vector potential formalism, which restores
translational invariance to the Hamiltonian at the expense of rendering the 
static field problem time-dependent~\cite{krieger86}.)

Let us now describe the minimization algorithm. 
We have chosen an iterative ``band-by-band'' conjugate-gradients 
method~\cite{payne92} in which each occupied state $u_{n \bf k}$
is updated in sequence, although other schemes may be equally suitable.
The many-electron state of interest violates inversion symmetry but not
time-reversal symmetry; the latter can be used, together with
any $\boldsymbol{\cal E}$-preserving point-group operations, in reducing the
BZ.  The few differences with respect to a normal ground-state
calculation stem from the $-\Omega\,{\bf P}_{\rm
mac}\cdot\boldsymbol{\cal E}$ term, as follows.
The gradient $| G_{n \bf k} \rangle=
\delta {\cal F}/\delta \langle u_{n \bf k}|$ becomes
\begin{eqnarray}
&&|G_{n \bf k} \rangle=\frac{f}{N_k}\, 
\hat{H}_{\rm KS}({\bf k}) | u_{n \bf k} 
\rangle
+ \frac{ife}{4\pi}\, \sum_{i=1}^3\, 
\frac{{\boldsymbol{\cal E}} \cdot {\bf a}_i}{N_{\perp}^{(i)}}\,
\times
\\ \nonumber
&&\quad
\sum_{m=1}^M
\Big[
| u_{m,{\bf k}_{+}^{(i)}} \rangle 
S^{-1}_{mn}({\bf k},{\bf k}_{+}^{(i)})-
| u_{m,{\bf k}_{-}^{(i)}} \rangle 
S^{-1}_{mn}({\bf k},{\bf k}_{-}^{(i)}) 
\Big],
\end{eqnarray}
where ${\bf k}_{\pm}^{(i)}={\bf k} \pm {\bf b}_i/N_i$ and use was made of 
Eq.~(88) 
of Ref.~\cite{nunes01}. By standard manipulations~\cite{payne92} this is
converted into a preconditioned conjugate-gradients search direction
$| D_{n \bf k} \rangle$ orthonormalized to the occupied manifold at ${\bf k}$.
The trial updated state is
${|\widetilde{u}_{n \bf k} \rangle}(\theta)=\cos \theta |u_{n \bf k} \rangle +
\sin \theta | D_{n \bf k} \rangle$. 
We search ${\cal F}(\theta)$ for a minimum, replace $|{u}_{n \bf k}\rangle$
by $|\widetilde{u}_{n \bf k}\rangle$, and go on to the next band.

\begin{figure}
\centerline{\epsfig{file=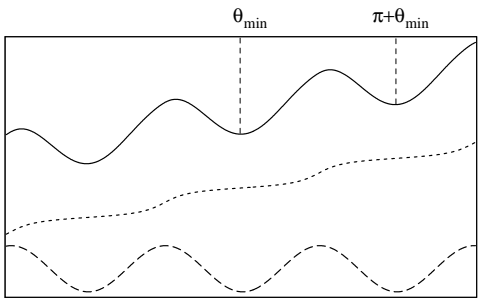,width=2.5in}}
\vspace{0.4cm}
\caption{The electric enthalpy ${\cal F}$
(solid line) and its components 
$E_{\rm KS}$ (dashed line) and
$-\Omega {\bf P}_{\rm mac} \cdot \boldsymbol{\cal E}$ (dotted line),
plotted as a function of the parameter $\theta$ that controls the update of a
polarized Bloch state in a conjugate-gradients step.}
\label{fig:theta}
\end{figure}

However, the behavior of ${\cal F}(\theta)$ is unconventional,
as is illustrated in Fig.~\ref{fig:theta}.
The $E_{\rm KS}(\theta)$ contribution (dashed line in Fig.~\ref{fig:theta})
is the usual one; it is periodic with period $\pi$ and has an 
amplitude
proportional to $E_{\rm gap}^{({\bf k})}/N_k$.  However,
$-\Omega {\bf P}_{\rm mac}(\theta) \cdot \boldsymbol{\cal E}$ (dotted line)
has a secular component arising from the fact that
${\bf P}_{\rm mac}$ changes by $1/N_\perp^{(i)}$ times a ``quantum
of polarization'' $\Delta{\bf P}=fe{\bf R}/\Omega$~\cite{ksv93},
where $\bf R$ is a (usually nonzero) lattice vector,
as $\theta\rightarrow\theta+\pi$. To understand this, consider one
phase $\beta={\rm Im}\ln\det S({\bf k},{\bf k}')$
contributing to Eq.~(\ref{berry-phase}).  Its $\theta$-dependence is
$\beta(\theta)={\rm Im}\ln\,(A\cos\theta+B\sin\theta)$, where
$A=\det S$ and $B=\det\widetilde{S}$ are complex constants
and $\widetilde S$ is obtained from $S$ by replacing the
$n$-th row with
$\langle D_{n\bf k}\vert u_{m{\bf k}'}\rangle$.  It is then
easy to see that $\beta(\theta)$ progresses by $\pm\pi$
as $\theta$ increases by $\pi$ (the sign depending on the sign of
${\rm Im}\,A^*B$).  In total there are six such contributions for
each $\bf k$; for $\boldsymbol{\cal E}$ along ${\bf b}_i$, the two involving
neighboring $\bf k'$ along the string direction $i$ contribute to 
an average slope $-fe\boldsymbol{\cal E}\cdot{\bf a}_i/N_\perp^{(i)}\pi$
of $-\Omega {\bf P}_{\rm mac}(\theta) \cdot \boldsymbol{\cal E}$
as a function of $\theta$, as shown by the dotted line in
Fig.~\ref{fig:theta}.
If $\cal E$ is not too large, we simply carry out the update by
stepping to the nearest local minimum of ${\cal F}(\theta)$.
(Local minima separated by $\pi$ are equivalent.)  However,
${\cal F}(\theta)$ {\it loses its minima when 
$\boldsymbol{\cal E}$ gets
too big.}  This occurs for $e\boldsymbol{\cal E}\cdot{\bf a}_i
\simeq E_{\rm gap}^{({\bf k})}/N_i$, which is precisely the heuristic
condition for the onset of Zener tunneling on a ring.

One possible concern with the present method is that it imposes a
minimum mesh spacing that can be used for a
given $\boldsymbol{\cal E}$. This
difficulty can be circumvented in practice by decomposing the fine
$k$ mesh into a set of sparser submeshes, computing the
Berry-phase terms on each submesh, and then averaging over
all of them.  Of course, $E_{\rm KS}$ can be computed
on the fine mesh.

We now turn to the computation of forces and stresses
at $\boldsymbol{\cal E}\not=0$.
According to the Hellmann-Feynman argument~\cite{payne92},
at a stationary point of ${\cal F}$ the force
${\bf F}_j \equiv - d{\cal F}/d{\bf r}_j$ becomes simply
$-\partial {\cal F}/\partial {\bf r}_j$, i.e.,
the implicit dependence via the wave functions can be dropped.
Eq.~(\ref{berry-phase}) has no such explicit dependence, and
so does not contribute to ${\bf F}_j$.  Thus, aside from the trivial ionic
core contribution $eZ_j\,\boldsymbol{\cal E}$, the force is given by
the standard $\boldsymbol{\cal E}$=0 Hellmann-Feynman expression
arising from $E_{\rm KS}$ alone.

As for the macroscopic stress, similar arguments
yield $\sigma_{\alpha \beta}=(1/\Omega)\partial {\cal F}/\partial 
\eta_{\alpha \beta}$, where 
$\boldsymbol{\eta}$ is a homogeneous strain.  However,
the electric boundary conditions under which the strain derivative is taken
must be specified carefully.  When the cell is deformed
according to ${\bf a}_i \rightarrow ({\bf 1}+\boldsymbol{\eta}){\bf a}_i$,
we can hold fixed either the macroscopic field $\boldsymbol{\cal E}$, or the
potential drop across each lattice vector,
$V_i=-\boldsymbol{\cal E}\cdot {\bf a}_i$. 
As the Berry phases $\varphi^{(i)}_{\rm el}$ do not depend explicitly on the
strain, it follows from the expression
\begin{equation}
\label{p_dot_e}
2\pi\,{\bf P}_{\rm el} \cdot {\boldsymbol{\cal E}}=
 \sum_{i=1}^3\, ({\boldsymbol{\cal E}} \cdot {\bf a}_i)
             ({\bf P}_{\rm el} \cdot 
{\bf b}_i) 
= \frac{fe}{\Omega}\, \sum_{i=1}^3\, ({\boldsymbol{\cal E}} \cdot 
{\bf a}_i)\, \varphi^{(i)}_{\rm el}
\end{equation}
that $\partial (\Omega {\bf P}_{\rm el}\cdot \boldsymbol{\cal E})/\partial
\eta_{\alpha \beta}=0$ when ${\bf V}$ is fixed
 (the same holds true for the ionic term, which can also
be recast in terms of a phase $\varphi^{(i)}_{\rm ion}$~\cite{dhv00}). As a 
result, the stresses in the two cases are related by
\begin{equation}
\label{stress}
\sigma_{\alpha\beta}^{(\boldsymbol{\cal E})}=
\sigma_{\alpha\beta}^{({\bf V})}-\frac{fe}{2\pi\Omega}\,
\sum_{i=1}^3\, {\cal E}_{\alpha}\,{(a_i)}_{\beta}\, (\varphi^{(i)}_{\rm el}+
\varphi^{(i)}_{\rm ion}),
\end{equation}
so that the 
pressures differ by
$({\bf P}_{\rm mac}\cdot \boldsymbol{\cal E})/3$.
The stress
$\boldsymbol{\sigma}^{({\bf V})}$ is given in terms of the polarized
Bloch states by the same expression as the stress in a zero-field 
ground-state calculation; it is a symmetric (torque-free) bulk quantity.
On the contrary, $\boldsymbol{\sigma}^{(\boldsymbol{\cal E})}$ generally
has an antisymmetric part, and moreover it depends on the choice of
branch cut when evaluating the multivalued bulk polarization.
(In the context of a finite crystallite, the torque and
extra stress in $\boldsymbol{\sigma}^{(\boldsymbol{\cal E})}$
can be regarded as arising from forces exerted on the polarization-induced
surface charges by a field that is held fixed in the 
``laboratory frame''~\cite{nelson79}.)
It is straightforward to show that
$c_{\alpha\beta\gamma}^{(\boldsymbol{\cal E})}= 
d \sigma_{\beta\gamma}^{(\boldsymbol{\cal E})}/
d{\cal E}_{\alpha}$ is the so-called ``improper'' piezoelectric 
tensor~\cite{dhv00}, whereas
\begin{equation}
\label{proper_piezo}
c_{\alpha\beta\gamma}^{({\bf V})}=\frac{d \sigma_{\beta\gamma}^{({\bf V})}}
{d{\cal E}_{\alpha}}=
-\, \sum_{i=1}^3\,
\frac{d \sigma_{\beta\gamma}^{({\bf V})}}{d V_i}\, a_{i,\alpha}
\end{equation}
is the ``proper'' tensor.

The scheme outlined above was first validated on a one-dimensional 
tight-binding model~\cite{nunes94}, where the results were 
found to agree with the results of the
Wannier-based approach~\cite{nunes94}. It was then implemented
in {\tt ABINIT}~\cite{abinit}, a fully self-consistent pseudopotential code.
To illustrate the utility of the method, we have calculated by  
finite-differences dielectric and piezoelectric constants of several
III-V semiconductors. That is, we increase $\boldsymbol{\cal E}$ in small
increments and compute the changes in the resulting forces, stresses, and
polarizations, with internal displacements and strain either clamped
or unclamped as appropriate. The Born effective charge is
$eZ^{*}_{j\alpha\beta}=dF_{j\beta}/d{\cal E}_{\alpha}$.
(Contrary to previous finite-difference approaches,
we compute it as the derivative of a force with respect to
$\boldsymbol{\cal E}$, not polarization with respect to displacement.)
The dielectric constant is 
$\epsilon_{\alpha\beta}=\delta_{\alpha\beta}+ \chi_{\alpha\beta}$, where
$\chi_{\alpha\beta}=(1/\epsilon_0)\, 
d{(P_{\rm mac})}_{\alpha}/d{\cal E}_{\beta}$
and $\epsilon_0$ is the vacuum permittivity. 
If the ions are kept fixed, this yields the electronic contribution
$\boldsymbol{\epsilon}_{\infty}$; if both electrons and ions are allowed to
relax in response to the field, the static dielectric constant
$\boldsymbol{\epsilon}_{\rm stat}$ is obtained. 
The quadratic susceptibility is 
$\chi^{(2)}_{\alpha \beta \gamma}=(2/\epsilon_0) 
d^2{(P_{\rm mac})}_{\alpha}/d{\cal E}_{\beta} d{\cal E}_{\gamma}$, and we
have computed it keeping the ions 
fixed. 
In the zinc blende structure, the only nonzero independent components of these
tensors are $Z^{*}_{11}$, 
$\epsilon_{11}$, and
$\chi^{(2)}_{123}$. The ``proper'' piezoelectric coefficient 
$c_{123}^{({\bf V})}$ is computed using
Eq.~(\ref{proper_piezo}), with both clamped ($\gamma_{14}^{(0)}$) and
unclamped ($\gamma_{14}$) ions.

 \begin{table}
 \begin{tabular}{lcccc}
 \hline \hline
 & GaAs & AlAs & GaP & AlP \\
 \tableline
 a (a.u.)     & 10.45 & 10.59 & 10.11 & 10.24 \\
              &(10.68)&(10.69)&(10.28)&(10.33)\\
 $Z^{*}_{\rm cation}$
              & 2.00 & 2.14 & 2.10 & 2.24 \\
              &(2.07)&(2.18)&(2.04)&(2.28)\\
 $\epsilon_{\infty}$
              & 11.9 & 9.6 & 9.4 & 8.1 \\
              &(10.9)&(8.2)&(9.0)&(7.5) \\
 $\epsilon_{\rm static}$ 
              & 13.5 & 11.5 & 11.2 & 10.2 \\
              &(13.2)&(10.1)&(11.1)&(9.8) \\
 $\chi^{(2)}$ (pm/V)
              & 134 & 64 & 66 & 39 \\
              &(166)&    &(74)& \\
 $(a^2/e)\,{\gamma}_{14}$
              & $-$0.40 & $-$0.10 & $-$0.25 & 0.05 \\
              &($-$0.32)&         &($-$0.18)& \\
 $(a^2/e)\,{\gamma}_{14}^{(0)}$
              & $-$1.42 & $-$1.40 & $-$1.35 & $-$1.31 \\
 \hline \hline
 \end{tabular}
 \caption{
 Computed dielectric and piezoelectric properties of III-V semiconductors. 
 Parentheses denote experimental data quoted in 
 Refs.~\cite{gironcoli89,dalcorso96,singh93,lucovsky71}.
 }
 \label{table:results} 
 \end{table}


The calculations were performed at the theoretical lattice constant $a$ using
an energy cutoff of 10~Ha.
We checked that our $k$ point sampling was very well converged at 
$16 \times 16 \times 16$
points in the full BZ. With this mesh spacing we find critical 
fields of the order of $10^7$\,V/cm, and the
finite-difference field step size was approximately 1/10 of
this value. We checked that our values for $Z^{*}$, ${\gamma}_{14}$, and
${\gamma}_{14}^{(0)}$ essentially coincide, 
for any given mesh of $k$ points, with those computed using the approach of
Refs.~\cite{ksv93,dhv00}.
We have also computed $Z^{*}$ and $\epsilon_{\infty}$  by treating the 
electric field via density-functional perturbation theory 
(DFPT)~\cite{gironcoli89}.
In the limit of a dense mesh the two approaches yield the same results; the 
discrepancies that occur for sparser meshes can be attributed to the different
ways in which derivatives with respect to ${\bf k}$ are handled in each
case~\cite{sai02}. Our
results for the piezoelectric coefficients and for $\chi^{(2)}$ are also in 
good agreement with experiment and with previous calculations using different 
methods~\cite{gironcoli89,dalcorso96,ksv93}.

All of the quantities reported in Table I could have been obtained using 
DFPT methods.  However, a considerable gain in
convenience is afforded by computing them using simple finite
differences of $\boldsymbol{\cal E}$.  For example, the calculation of
$\chi^{(2)}$ by DFPT is quite tedious and requires
a special-purpose program, while $\chi^{(n)}$ of arbitrary order
are easily computed using the present approach.

To summarize, we have presented a practical first-principles
scheme for computing the electronic structure of insulators under a finite
dc bias. 
The algorithm is ideally suited for implementation in a standard
electronic structure code and its computational cost is
comparable.  Dielectric polarization, forces and stresses are
straightforwardly obtained as byproducts of the calculation.
The extension of this approach to time-dependent fields
will be discussed in a future communication.

This work was supported by NSF Grant DMR-9981193.
We thank Ralph Gebauer for stimulating discussions.


\begin{thebibliography}{0}


\bibitem{kane59} E. O. Kane, J. Phys. Chem. Solids {\bf 12}, 181 (1959).

\bibitem{wannier60} G. H. Wannier, Phys. Rev. {\bf 117}, 432 (1960).

\bibitem{nenciu91} 
G. Nenciu, Rev. Mod. Phys. {\bf 63}, 91 (1991).

\bibitem{gironcoli89} S. de Gironcoli, S. Baroni, and R. Resta,
Phys. Rev. Lett. {\bf 62}, 2853 (1989);
P. Giannozzi {\it et al.}, Phys. Rev. B {\bf 43},
7231 (1991);
X. Gonze, Phys. Rev. B {\bf 55}, 10\,337  (1997).

\bibitem{dalcorso96} 
A. Dal Corso, F. Mauri, and A. Rubio, Phys. Rev. B 
{\bf 53}, 15\,638 (1996).

\bibitem{putrino00} A. Putrino, D. Sebastiani, and M. Parrinello,
J. Chem. Phys. {\bf 113}, 7102 (2000).

\bibitem{nunes01} R. W. Nunes and X. Gonze, Phys. Rev. B {\bf 63}, 155107
(2001).

\bibitem{sai02} N. Sai, K. Rabe, and D. Vanderbilt, cond-mat/0205442.   

\bibitem{zener34} C. Zener, Proc. Roy. Soc. (London) {\bf 145}, 523 (1934).

\bibitem{foot-atoms} Charge leakage also occurs in
systems with surfaces (field-induced autoionization).  
However, it is easily avoided by using a standard localized-orbital basis
that spans only the region where the system is located, which
prevents tunneling into distant regions. In solid-state physics,
on the other hand, such distant regions must be covered by basis
functions because they are part of the sample.

\bibitem{kunc83} K. Kunc and R. Resta, Phys. Rev. Lett. {\bf 51}, 686 (1983).

\bibitem{ksv93} R. D. King-Smith and D. Vanderbilt, Phys. Rev. B {\bf 47},
1651 (1993); D. Vanderbilt and R. D. King-Smith, Phys. Rev. B {\bf 48},
4442 (1993). 
Although the original derivation assumed
shorted boundary conditions, we have shown that the same formula remains valid
for $\boldsymbol{\cal E}\not=0$~\cite{unpublished}.

\bibitem{nunes94} R. W. Nunes and D. Vanderbilt, Phys. Rev. Lett. {\bf 73},
712 (1994).

\bibitem{fernandez98} P. Fernandez, A. Dal Corso, and A. Baldereschi, 
Phys. Rev. B {\bf 58}, R7480 (1998).

\bibitem{unpublished} I. Souza, J. \'I\~niguez, and D. Vanderbilt
(unpublished).

\bibitem{kohn68} W. Kohn, Phys. Rev. {\bf 133}, A171 (1964).

\bibitem{souza00} I. Souza, T. Wilkens, and R. M. Martin, Phys. Rev. B 
{\bf 62}, 1666 (2000).

\bibitem{krieger86} J. B. Krieger and G. J. Iafrate, Phys. Rev. B {\bf 33},
5494 (1986); R. Gebauer and R. Car (unpublished).

\bibitem{payne92} M. C. Payne {\it et al.}, Rev. Mod. Phys. {\bf 64}, 1045
(1992).

\bibitem{dhv00} D. Vanderbilt, J. Phys. Chem. Solids {\bf 61}, 147 (2000).

\bibitem{nelson79} D. F. Nelson, {\em Electric, Optic, and Acoustic
Interactions in Dielectrics} (Wiley, New York, 1979).

\bibitem{abinit} The {\tt ABINIT} code is a common project of the Universit\'e 
Catholique de Louvain, Corning Incorporated, and other contributors 
(URL http://www.abinit.org).

\bibitem{singh93} J. Singh, {\em Physics of Semiconductors and Their
Heterostructures} (McGraw-Hill, New York, 1993).

\bibitem{lucovsky71} G. Lucovsky, R. M. Martin, and E. Burstein,
Phys. Rev. B {\bf 4}, 1367 (1971).

\end{thebibliography}
\end{document}